\documentclass[onecolumn,trackchanges]{aastex6}

\newcommand{\new}[1]{{#1}}   

\newcommand{\tff}{t_{\mbox{\scriptsize{\em ff}}}}

\shorttitle{Molecular Cloud Formation and Destruction}
\shortauthors{Mac Low et. al.}

\begin{document}

\title{Fast Molecular Cloud Destruction Requires Fast Cloud Formation}

\author{Mordecai-Mark Mac Low\altaffilmark{1,2}, Andreas Burkert\altaffilmark{3,4}, Juan C. Ib\'a\~nez-Mej\'{\i}a \altaffilmark{4,5}}

\altaffiltext{1}{American Museum of Natural History, 79th Street at Central Park West, New York, NY 10024, USA }
\altaffiltext{2}{Universit\"at Heidelberg, Zentrum f\"ur Astronomie Heidelberg, Institut f\"ur Theoretische Astrophysik, 69120 Heidelberg, Germany}
\altaffiltext{3}{Universit\"ats Sternwarte M\"unchen, Ludwigs-Maximilian-Universit\"at, 81679 M\"unchen, Germany}
\altaffiltext{4}{Max-Planck-Institut f\"ur Extraterrestrische Physik, 85748 Garching bei M\"unchen, Germany}
\altaffiltext{5}{I. Physikalisches Institut, Universit\"at zu K\"oln, 50937 K\"oln, Germany}
\email{mordecai@amnh.org, burkert@usm.lmu.de, ibanez@ph1.uni-koeln.de}

\begin{abstract}

A large fraction of the gas in the Galaxy is 
cold, dense, and molecular.  If all this gas collapsed under the influence of gravity and formed stars in a local free-fall time, the star formation rate in the Galaxy would exceed that observed by more than an order of magnitude. Other star-forming galaxies behave similarly. Yet observations and simulations both suggest that the molecular gas is indeed gravitationally collapsing, albeit hierarchically. Prompt stellar feedback offers a potential solution to the low observed star formation rate if it quickly disrupts star-forming clouds during gravitational collapse. However, this requires that molecular clouds must be short-lived objects, raising the question of how so much gas can be observed in the molecular phase.  This can occur
only
if molecular clouds form as quickly as they are destroyed, maintaining a global equilibrium fraction of dense gas.  We therefore examine cloud formation timescales.  We first demonstrate that supernova and superbubble sweeping cannot produce dense gas at the rate required to match the cloud destruction rate.  On the other hand, Toomre gravitational instability can reach  the required production rate.  
We thus argue that, although dense, star-forming gas may last only around a single global free-fall time, the dense gas in star-forming galaxies can globally exist in a state of dynamic equilibrium between formation by gravitational instability, and disruption by stellar feedback.  At redshift $z \gtrsim 2$, the Toomre instability timescale decreases, resulting in a prediction of higher molecular gas fractions at early times, in agreement with observations.

\end{abstract}

\maketitle

\section{Introduction}
\label{sec:Introduction}

Beginning in 1970, radio observations of CO \citep{wilson1970} revealed the extent of the dense, cold, molecular gas in the Galaxy.  \citet{ferriere2001} concludes that \new{the molecular mass} $M_m \simeq 2\times 10^9$~M$_{\odot}$,
over 20\% of the total \new{gas} mass.
The supersonic velocity dispersions seen in the cold gas were quickly interpreted to be driven by gravitational collapse \citep{goldreich1974}, as the inferred temperatures and densities of the gas suggested that much of it is gravitationally unstable. The objection was raised by \citet{zuckerman1974} that if that much gas formed stars within a free-fall time
\begin{equation}
\tff = (3 \pi / 32 G \rho)^{1/2} \simeq (3.7 \mbox{ Myr}) (n /100\mbox{ cm}^{-3})^{-1/2},
\end{equation}
where the mean mass per particle $\mu_m = \rho/n = 3.32 \times 10^{-24}$~g for molecular gas with a helium to hydrogen atomic ratio of 0.1 (neglecting the few percent by mass contributed by heavy elements), then the star formation rate would be 
\begin{equation} \dot{M}_* = M_m / \tff \simeq (540 \mbox{ M}_{\odot} \mbox{ yr}^{-1})  (n /100\mbox{ cm}^{-3})^{1/2},
\end{equation} 
more than \new{two orders} of magnitude higher than the observed value of 1--2~M$_{\odot}$~yr$^{-1}$ \citep[][and references therein]{licquia2015}.

This paradox led to intensive exploration of equilibrium models of molecular clouds, with support against gravity provided by some mechanism.  Proposed support mechanisms have included magnetic fields until the onset of ambipolar diffusion \citep{mouschovias1977}, or turbulence driven by protostellar jets \citep{nakamura2007} or ionization 
\citep{matzner2002}.  However observations of magnetic field strengths and configurations  \citep{crutcher1999} and column density contrasts \citep{nakano1998} have ruled out magnetic equilibrium, while the demonstration by \citet{brunt2009} that observed molecular cloud motions have the most power at the largest available scales argues against internal sources of turbulent driving.

Increasing amounts of evidence support the original interpretation of \citet{goldreich1974} that molecular clouds are indeed gravitationally collapsing.  The quick formation of stars across the Taurus cloud was used by \citet{ballesteros-paredes1999} to argue that only global gravitational collapse could synchronize star formation along the entire length of the cloud.  \citet{heyer2009} demonstrated that clouds observed with multiple tracers that could follow a broad range of surface density $\Sigma$ have a dependence $\Sigma \propto \sigma^2 / r^{1/2}$ on their velocity dispersion $\sigma$ and radius $r$, as expected for either free-fall or virial equilibrium.  \citet{ballesteros-paredes2011} argued that this indeed represents free-fall, noting that the difference in expected velocity magnitude is only a factor of $2^{1/2}$ larger than would be expected for virial equilibrium.
Numerical simulations by \citet{ibanez-mejia2016} of the supernova-driven interstellar medium showed that external turbulent driving could not generate the observed velocity dispersion relations, but the action of self-gravity does reproduce those relations.
(\citealt{dobbs2012a} argue for a similar conclusion in many cases, while \citealt{padoan2016} draw the opposite conclusion from models with a higher supernova driving rate.)  

The observed inefficient star formation then remains a puzzle. 
Presumably, some form of stellar feedback 
disperses or destroys molecular clouds before they 
    convert a large fraction of their mass into stars 
\citep[e.g.][]{elmegreen2000}. External feedback from superbubbles and field supernovae appears insufficient to perform this task \citep{ibanez-mejia2016}, so the feedback must come from within the clouds.  Ionization \citep{matzner2002} and radiation pressure \citep{murray2010} have been suggested for the largest clouds, while protostellar outflows may also contribute in smaller clouds 
\citep{matzner2002,nakamura2007}.

Does feedback simply disperse the self-gravitating dense regions in molecular clouds, or does it destroy clouds entirely?  The low ages of molecular cloud cores ($< 2$~Myr) derived from chemical 
abundances \citep[e.g.][]{ohishi1992,van-dishoeck1998} argue for destruction and reformation. If clouds are destroyed entirely, however,
     then new molecular gas must form quickly enough to explain the observed molecular fraction in galaxies.
To quantify this argument, we can take cloud lifetimes to certainly be $\tau_c \ge 10$~Myr. 
    If we assume all molecular gas goes through the same formation and destruction cycle, we can balance
    the molecular cloud destruction rate against the cloud formation rate. 
If we take the minimum cloud lifetime $\tau_{c}=10$~Myr, the molecular gas formation rate needs to be no more than 
\begin{equation} \dot{M}_{m} = M_{m}/\tau_{c}=200\mbox{ M}_{\odot}\mbox{ yr}^{-1}.  
\label{eq:rate}
\end{equation} 
In this paper, we argue that clouds can indeed form as quickly as they must be destroyed, so that the continuity equation can be satisfied even if feedback entirely dissociates the molecular gas in star forming regions. 
\new{\citet{semenov2017} perform a more generalized analysis of the continuity equation to reach similar conclusions on how star formation proceeds using both analytic arguments and comparisons to a numerical model of a full galaxy.}

In Sections 2 and 3 we examine whether sufficient gas can be swept up by either supernova remnants and superbubbles or \citet{toomre1964} gravitational instability to form new clouds at the rate that they must be destroyed to maintain the observed molecular gas fraction in the Galaxy. We find that the combination of supernova and superbubble sweeping is insufficient, but that gravitational instability can indeed form dense gas at the required rate.  This is consistent with observed clouds being in a state of hierarchical gravitational collapse.  In Section 4, we explore the implication for galaxies over cosmic time, showing that the expected shortening of the instability time scale in young galaxies with high surface densities should then lead to higher dense gas fractions, in agreement with observations.  Finally, in Section 5, we summarize.

\section{Supernovae}

Can supernovae sweep up dense gas fast enough to form clouds at the required rate?  If we take a canonical time between Milky Way supernovae of $\tau_{\rm SN} = 40$~yr \citep{tammann1994}, that would require that each supernova sweep up 
\begin{equation} M_{\rm SN} = \dot{M}_m \tau_{\rm SN} = 8 \times 10^3\mbox{ M}_{\odot}.
\label{eq:SNrate}
\end{equation}  
We therefore attempt to estimate how much gas supernovae or superbubbles can compress to the required densities.  We note that the molecule formation time \citep{hollenbach1971}
\begin{equation}\tau_f= 1 \mbox{ Gyr} (n / 1 \mbox{ cm}^{-3})^{-1}. \end{equation} Thus densities of order 
$1000$~cm$^{-3}$ must be reached to form molecules in a time comparable to the lifetime of the supernova shell.

    We start by making the optimistic assumption for sweeping up clouds that every supernova  explodes in gas of the mean density in the disk,
with number density $n = 1$~cm$^{-3}$.  
   This ignores concentration of gas in dense clouds.  We further assume
 that they sweep gas up into a shell until they reach a velocity 
similar to the background velocity dispersion of 10~km~s$^{-1}$.

However, not all the atomic gas swept up by the shock will be dense enough to become molecular.  A plausible requirement for interstellar gas with an average density of 1~cm$^{-3}$ and temperature of around $10^4$~K to reach the required density of $\sim 1000$~cm$^{-3}$ is that the supernova shock compress the gas by a factor of ten in an isothermal shock, followed by cooling from $T = 10^4$~K to $10^2$~K, producing another factor of \new{hundred} compression.  Since isothermal shocks compress the shocked gas as the square of the Mach number ${\cal M}^2$, this requires that the shock be traveling at ${\cal M} = 10^{0.5}$.  If we take the background sound speed to be 10~km~s$^{-1}$, this implies a minimum velocity of $v_{cr} = 10^{1.5}$~km~s$^{-1}$ for molecule formation. 

\subsection{Adiabatic}
A supernova blast wave in the most optimistic case with interior radiative cooling neglected will expand to a radius of 
\begin{equation} 
R_{\rm SN} = (2.026 E / \rho)^{1/5} t^{2/5} \simeq (81\mbox{ pc}) 
\left(\frac{E}{10^{51}\mbox{ erg s}}\right)^{1/5}
\left(\frac{n}{1\mbox{ cm}^{-3}}\right)^{-1/5}
\left(\frac{t}{1 \mbox{ Myr}}\right)^{2/5}
\end{equation}
\citep{ostriker1988}, where we have assumed the mean mass per particle $\mu_a = \rho/n = 2.11 \times 10^{-24}$~g appropriate for neutral atomic gas, again with helium to hydrogen atomic ratio of 0.1.
The velocity of the remnant is then by differentiation
\begin{equation}  \label{eqn:vsn}
v_{\rm SN} =  (2.075 \times 10^{-2} E / \rho)^{1/5} t^{-3/5} \simeq (32 \mbox{ km s}^{-1})
\left(\frac{E}{10^{51}\mbox{ erg s}}\right)^{1/5}
\left(\frac{n}{1\mbox{ cm}^{-3}}\right)^{-1/5}
\left(\frac{t}{1 \mbox{ Myr}}\right)^{-3/5}
\end{equation}
   Thus, under the stated assumptions, we find that the expanding shell will drop below the critical velocity $v_{cr}$ at a time $t = 1~$Myr, having swept up a mass $M_{\rm SN} = 5.2 \times 10^4$~M$_{\odot}$, still comfortably above the required amount \new{given in Equation~\ref{eq:SNrate}}.

\subsection{Radiatively Cooled}

However, this estimate neglects radiative cooling of the supernova remnant interior. \citet{cioffi1988} derived an offset power-law for expansion of a radiatively cooled remnant into a uniform medium, 
\begin{eqnarray}
R_{\rm SN} &= & R_{{\rm PDS}} \left(\frac43 \frac{t}{t_{{\rm PDS}}} - \frac13\right)^{3/10}\\
v_{\rm SN} & = &v_{{\rm PDS}} \left(\frac43 \frac{t}{t_{{\rm PDS}}} - \frac13\right)^{-7/10},
\end{eqnarray}
where
\begin{eqnarray}
R_{{\rm PDS}} & = & (14.0\mbox{ pc}) Z^{1/7} \left(\frac{E}{10^{51}\mbox{ erg}}\right)^{2/7} \left(\frac{n}{1\mbox{ cm}^{-3}}\right)^{-3/7},  \\
v_{{\rm PDS}} & = & (413\mbox{ km s}) Z^{3/14} \left(\frac{E}{10^{51}\mbox{ erg}}\right)^{1/14} \left(\frac{n}{1\mbox{ cm}^{-3}}\right)^{1/7}, \\
t_{{\rm PDS}} & = & (1.28 \times 10^4 \mbox{ yr}) Z^{5/14} \left(\frac{E}{10^{51}\mbox{ erg}}\right)^{3/14} \left(\frac{n}{1\mbox{ cm}^{-3}}\right)^{-4/7},
\end{eqnarray}
and $Z$ is the metallicity relative to solar. Using this model, with the same density as above of $n = 1$~cm$^{-3}$, the velocity reaches $v_{cr} = 10^{1.5}$~km~s$^{-1}$ at a time of 0.387~Myr and a radius of 42.5 pc, sweeping up a mass of $M_{\rm SN} = 1.02 \times 10^4 \mbox{ M}_{\odot}$, now just barely exceeding the required value of $8 \times 10^3 \mbox{ M}_{\odot}$ \new{(Eq.~[\ref{eq:SNrate}])}.

\subsection{Clustering}

Supernova explosion locations are not, however, random, as
   roughly 60\% of core-collapse 
supernovae occur within OB associations \citep{cowie1979,avillez2000}.  This clustering results in the formation of superbubbles that can vent substantial portions of their energy into the galactic halo from the disk.
   As a result, core-collapse supernovae do not typically encounter gas 
   close to the average disk density of $n=1$~cm$^{-3}$, but instead
   explode within the low-density interiors of superbubbles.  Note that, although stellar winds are important during the 
   first few megayears of expansion, their integrated energy is only $\sim 10$\% 
   of the total supernova energy \citep[cf.\ right panel of Figure 2 in][]{shull1995}.   The radius of a 
   superbubble containing $N_{\rm SN}$ supernovae exploding at times $t \leq t_{\rm SB}$, 
   \begin{equation}
      R_s = \left(\frac{125}{154 \pi}\right)^{1/5} 
      \left(\frac{N_{\rm SN} E}{t_{\rm SB}}\right)^{1/5} \rho^{-1/5} t^{3/5} \label{Rs}
   \end{equation}
 \new{  \citep[e.g.][]{ostriker1988}} depends quite sublinearly on the total energy input 
   $N_{\rm SN} E$.
   Therefore, superbubbles sweep up mass  
   \begin{equation} M_s = (4/3) \pi \rho R_s^3 \label{Ms}\end{equation}
   far less efficiently than isolated supernova remnants.  

   We can compute the final 
   swept-up mass as a function of the critical velocity by computing the time 
   $t_{cr}$ at which the superbubble expansion velocity 
   \begin{equation} v_s =  \frac35 \frac{R_s}{t} \label{vs} \end{equation} 
   drops to $v_{cr}$, and substituting back into 
   Equations~(\ref{Rs}) and~(\ref{Ms}).  Dividing by $N_{\rm SN}$ 
   allows direct comparison to the required value of mass swept up per supernova 
   explosion: 
   \begin{equation}
    \frac{M_s}{N_{\rm SN}} \simeq (476 \mbox{ M}_{\odot}) 
    \left(\frac{N_{\rm SN}}{40}\right)^{1/2} 
    \left(\frac{E}{10^{51}\mbox{ erg}}\right)^{3/2} 
    \left(\frac{t_{\rm SB}}{40 \mbox{ Myr}}\right)^{-3/2}
    \left(\frac{n}{1\mbox{ cm}^{-3}}\right)^{-1/2}
    \left(\frac{v_{cr}}{10^{1.5} \mbox{ km s}^{-1}}\right)^{-9/2}. \label{MperSN}
   \end{equation}
   The scale of $t_{\rm SB}$ was chosen to be the typical lifetime of an
   8~M$_{\odot}$ star, likely the smallest star to explode as a supernova \citep{shull1995}. The required value of the mass swept up per supernova exceeds by more than an order of magnitude that produced by the small cluster chosen here for scaling.

However, the mass per supernova does depend on $N_{\rm SN}^{1/2}$, so larger clusters could perhaps sweep up sufficient mass.  This is ultimately limited by the thickness of the galactic disk, since a superbubble growing much larger than a scale height $H$ will blow out of the disk, venting the energy of all further supernovae to the halo, and only expanding within the disk plane at velocities far below $v_{cr}$.  The largest amount of mass swept up per supernova is by a superbubble whose velocity just drops to $v_{cr}$ at a scale height.  We can compute $N_{\rm SN}$ for such a superbubble by first computing the time $t_H$ at which a superbubble reaches $R_s = H$, and then substituting those values into Equation~(\ref{vs}) for the velocity. Setting $v_s = v_{cr}$ and solving for $N_{\rm SN}$, we find \begin{equation}
    N_{\rm SN, max} = 108  \left(\frac{v_{cr}}{10^{1.5} \mbox{ km s}^{-1}}\right)^3
    \left(\frac{E}{10^{51}\mbox{ erg}}\right)^{-1} 
    \left(\frac{t_{\rm SB}}{40 \mbox{ Myr}}\right)
    \left(\frac{n}{1\mbox{ cm}^{-3}}\right)
    \left(\frac{H}{200 \mbox{ pc}^{-3}}\right)^2. \label{NSNmax}
\end{equation}
Substituting this maximal value of $N_{\rm SN}$ into Equation~(\ref{MperSN}), we find a swept up mass per supernova of 782~M$_{\odot}$ for the typical parameters chosen for scaling, still a full order of magnitude below the required value of $8 \times 10^3 \mbox{ M}_{\odot}$ \new{(Eq.~[\ref{eq:SNrate}])}.

\subsection{Other Evidence}
Indeed, both observations and numerical simulations support the idea that neither supernovae nor superbubbles can sweep up enough mass to explain the required formation rate of molecular clouds.  
\citet{dawson2013} showed that only $\sim10$\% of the molecular mass in the Large Magellanic Cloud was formed as a result of sweeping by supergiant shells.  
Simulations of supernova-driven turbulence in a representative section of a stratified galactic disk by \citet{joung2006} came to a similar conclusion. Neglecting the action of self-gravity on the gas, they found that only 10\% as much gas was swept up into regions able to gravitationally collapse as would have been required to maintain the star formation rate implied by the input supernova rate.

\section{Toomre Instability}

Therefore, we now turn to large-scale gravitational instability in disks \citep{safronov1960,toomre1964,goldreich1965}, which offers an alternative mechanism to assemble gas dense enough to form molecules. Galactic dynamics appears to be driven by such instabilities, which at the largest scales drive spiral arm formation \citep{agertz2009,wada2011}.
When these instabilities grow strongly enough because of gas infall or galaxy mergers they lead to the formation of giant clumps \citep{li2005a} in what is called violent disk instability 
\citep{dekel2009,agertz2009}. (Those clumps may well themselves collapse into clusters of small clumps as shown by \citealt{behrendt2016}.)  That molecular clouds can form more quickly by gravitational instability than collisional agglomeration was shown by \citet{elmegreen1990}, despite their apparently having the same dependences on density and velocity dispersion.

The Toomre instability criterion for gaseous disks \citep{goldreich1965} is
\begin{equation}
Q_g = \kappa \sigma_g / (\pi G \Sigma_g)< 1,
\end{equation}
where $\Sigma_g$ is the surface density of the gas, $\sigma_g$ is its velocity dispersion including both thermal and nonthermal components, $G$ is the gravitational constant, and $\kappa^2 =-4 \omega B$ is the square of the epicyclic frequency, with Oort's constant 
\begin{equation}
B = - \frac12 \left[ \omega + \frac{\partial (\omega r)}{\partial r} \right]. 
\end{equation}
In a galaxy with a flat rotation curve, the second term in $B$ can be neglected.  In this case, $\kappa = \omega \sqrt{2}$, so that $ Q_g = \omega \sigma_g \sqrt{2}/ ( \pi G \Sigma_g) $. The timescale for growth of a perturbation in an isothermal gas disk can be expressed in terms of $Q_g$ as \citep{wang1994} 
\begin{equation}
\tau_g = \frac{Q_g}{\kappa (1 - Q_g^2)^{1/2}} = \kappa^{-1} f(Q). 
\end{equation} 
At the solar circle, with $r = 8$~kpc, the rotational velocity $\omega r = 220$~km~s$^{-1}$, dropping only slowly to 170~km~s$^{-1}$ at a Galactic radius of 60~kpc \citep{xue2008}.  
Thus, the epicyclic frequency $\kappa = 1.26 \times 10^{-15}$~s$^{-1}$, and the growth time $\tau_g = f(Q) \kappa^{-1}$, where $\kappa^{-1} \simeq 25$~Myr, and $f(Q)$ is a function of $Q$ with a value of order unity.
If the diffuse gas with mass $M_d \sim 4 M_m$ \citep{ferriere2001} forms into molecular clouds on the Toomre timescale, then the molecular gas formation rate would be 
\begin{equation}
\dot{M}_m = M_d / \tau_g = 4 M_m / \tau_g 
\simeq 320 \mbox{ M}_{\odot} \mbox{ yr}^{-1}.
\end{equation} 
If the molecular gas is being dispersed on a timescale of order 10~Myr, this formation rate can replenish it at 
\new{more than the required rate given in Equation~(\ref{eq:rate})|}. 

\section{Variation with Redshift}

\new{We now consider how the timescale for molecular gas formation by gravitational instability varies with redshift.} The Toomre timescale $\tau_g \propto \omega^{-1}$. The disk angular velocity of a galaxy $\omega$ can be characterized by its value at the exponential scale length of the disk $\omega(R_d) = V_c/R_d$, where $V_c$ is the roughly constant circular velocity of the disk and halo.  

The disk scale length can be expressed as \citet{mo1998} to be
\begin{equation}
R_d = 2^{-1/2} (j_d / m_d) \lambda R_{200}, \label{eq:Rd}
\end{equation}
where $j_d$ and $m_d$ are the fractions of total halo angular momentum and mass in the disk, $\lambda$ the spin parameter of the halo, which compares the actual angular speed of the halo to that required for centrifugal support, and $R_{200}$ is the radius at a density of 200 times the critical closure density of the universe, roughly the halo boundary \citep{bertschinger1985}. However, halos at higher redshift have smaller radii \citep{mo1998}
\begin{equation} R_{200} = V_c / [10 H(z)],   \label{eq:R200}
\end{equation}
where the Hubble parameter as a function of redshift
\begin{equation} H(z) = H_0 [\Omega_{\Lambda,0} + (1-\Omega_{\Lambda,0} - \Omega_{M,0})(1+z^2) +   \Omega_{M,0} (1 + z)^3]^{1/2}, \end{equation}
The parameters $H_0$, $ \Omega_{M,0}$, and $\Omega_{\Lambda,0}$ are the present day values of the Hubble parameter, and the total matter density and cosmological constant normalized by the critical density for closure.  

The relationship between disk and halo radius depends on $\lambda$, $j_d$, and $m_d$, none of which appear to have strong redshift dependence. The spin parameter $\lambda$ of disks with $0.8 < z < 2.6$ has been measured to have a log-normal distribution around $\lambda = 0.035$ \citep{burkert2016} with a dispersion of only 0.2 in the log and no redshift dependence.  This is entirely consistent with the value predicted for dark matter halos by simulations \citep{bullock2001,hetznecker2006}. 
For galaxies on the star formation main sequence \citet{burkert2016} find typical values of $m_d=0.05$ and $j_d = 1$, independent of redshift.
The agreement between disk and halo angular speed furthermore implies that $j_d$ is roughly constant, a conclusion also supported by the success of models making this assumption \citep[e.g.][]{mo1998},
and consistent with modern numerical simulation results \citep[e.g.][]{ubler2014}. It is worth noting clearly that no fundamental reason for this agreement has yet been identified, as it occurs as a result of competition between the different angular momentum transfer mechanisms.  The disk mass fraction $m_d$ appears to be mass dependent, but \citet{mo1998} assume no variation with redshift.

We can therefore conclude that the redshift dependence of the Toomre timescale $\tau_g$ for a halo of given mass is contained in the halo radius variation given by equation~(\ref{eq:R200}). Substituting from that equation into equation~(\ref{eq:Rd}) gives
\begin{equation}
R_d  =  2^{-1/2} \left( \frac{j_d}{m_d}\right) \lambda \frac{V_c}{10 H(z)}.
\end{equation}    
Thus, the angular velocity of the disk 
\begin{equation}
\omega(R_d) = 2^{1/2} \left( \frac{m_d}{j_d}\right) \frac{10 H(z)}{\lambda}.
\end{equation}
As the Hubble parameter increases at higher redshift, the Toomre timescale thus declines as $\tau_g \propto 1/H(z)$.

At redshift $z \sim 2$, the formation timescale for molecular clouds from gravitational instability $\tau_g$ has already declined by a factor of about three because of this effect. 
       The destruction timescale might remain constant, 
as might be expected if it were primarily determined by small-scale feedback processes, 
        or it might increase as gravitational instability occurs more strongly.  
        Either way,
one reaches the conclusion that there should have been a higher fraction of molecular gas at earlier times. 

\new{Observations of high-redshift galaxies in CO show molecular gas masses that are 0.15--0.5 of the total baryonic mass at $z = 1$ and 0.3--0.8 at $z = 2$ \citep{tacconi2010}.  These high ratios leave little possibility for substantial masses of atomic gas without violating dynamical constraints on the total baryonic mass \citep{daddi2010}, thus implying high molecular gas fractions even in the absence of explicit measurements of atomic gas, consistent with our conclusion. }
 
\section{Summary}

We have considered the implications of the rapid formation of molecular gas in gravitationally collapsing molecular clouds.  To prevent the rapid formation of all available gas into stars, on timescales much shorter than observed gas depletion times, clouds must be destroyed rapidly after they begin to collapse.  This then raises the question of how the observed molecular gas fraction in galactic disks is maintained.  We have shown here 
   that isolated supernovae in average density gas can just barely sweep up enough molecular mass, but that the correlated locations of most core-collapse supernovae reduces their efficiency by an order of magnitude or more. Summing over the full population suggests that supernovae cannot explain observed molecular mass fractions.  On the other hand, 
Toomre gravitational instability
    can indeed assemble dense gas quickly enough
to maintain observed molecular gas fractions. Toomre instability is expected to become even stronger at higher redshift, predicting higher molecular fractions at high redshifts as a general consequence, in agreement with observations.  
   Thus, a fast cycle of gas collapsing to star-forming densities by gravitational instability followed by quick dispersal from stellar feedback appears consistent with observations of molecular gas.

\acknowledgments{
  M-MML and JCI-M were partially supported by NSF grant AST11-09395.  M-MML was additionally supported by the Alexander von Humboldt-Stiftung. JCI-M was additionally supported by the DFG Priority Programme 1573. AB acknowledges support by the Max-Planck-Institut f\"ur Extraterrestrische Physik. \new{We thank the referee for a careful review.}
}

\bibliographystyle{aasjournal}



\end{document}